\documentclass[proceedings, preprint]{rmaa}

\usepackage{paralist}

\usepackage{psfrag,color}
\usepackage{xcolor}
\usepackage[T1]{fontenc}

\newcommand{\empire}{\textsc{EMPIRE}}
\newcommand{\darkred}{red!70!black}
\newcommand{\darkblue}{cyan!60!black}
\newcommand{\lightblue}{cyan}

\newcommand{\rmxaa}{Rev. Mex. Astron. Astrofis.}

\SetYear{2023}
\SetConfTitle{XVII RRLA-LARIM}

\title{The co-evolution between galaxies and dark matter halos} 

\author{
  Aldo Rodriguez-Puebla\altaffilmark{1}}

\altaffiltext{1}{Universidad Nacional Aut\'onoma de M\'exico, Instituto de Astronom\'ia, A.P. 70-264, 04510, CDMX, M\'exico (apuebla@astro.unam.mx).}

\shortauthor{Rodriguez-Puebla}
\shorttitle{Galaxy-Halo co-evolution}

\listofauthors{Rodriguez-Puebla}
\indexauthor{Aldo Rodriguez-Puebla}

\abstract{The current cosmological paradigm asserts that dark matter halos provide the gravitational scaffolding for galaxy formation through a combination of hierarchical structure formation and non-linear local (g)astrophysical processes. This close relationship, known as the galaxy-halo connection, suggests that the growth and assembly of dark matter halos impact the properties of galaxies. While the stellar mass of galaxies correlates strongly with the mass of their dark matter halos, it is important to note that the galaxy-halo connection encompasses a broader distribution of galaxy and halo properties. This distribution can be constrained using data from astronomical observations and cosmological $N$-body simulations, a technique known as semi-empirical modeling. By operating at the intersection of observational data and the cosmological structure formation model, the semi-empirical modeling provides valuable insights into galaxy formation and evolution from a cosmological perspective. In this proceeding, we utilize a new sEM-emPIRical modEl, \empire, to explore the star formation history (SFH) of central galaxies across cosmic epochs, spanning from dwarfs to massive ellipticals. \empire\ aims to constrain the multivariate distribution that links galaxy and halo properties. Our findings reveal distinct growth stages for progenitors of central massive galaxies. Evidence suggests that cold streams played a significant role in sustaining star formation at higher redshifts, while virial shock heating became more prominent at lower redshifts. We observe that the maximum star formation efficiency occurs at approximately a factor of $\sim1.5-2$ below $M_{\rm vir \; shocks}$ for $z\lesssim1$. Furthermore, at higher redshifts, $z>1$, this peak tends towards higher masses, approximately $M_{\rm vir}\sim 2\times 10^{12} M_{\odot}$. Notably, at redshifts higher than $z\sim2$, the peak of star formation efficiency aligns comfortably within the region characterized by cold streams.
}

\addkeyword{galaxies: evolution}
\addkeyword{galaxies: halos}
\addkeyword{methods: statistical}

\begin{document}
\maketitle

\section{ \textbf{{\color{\darkblue} Introduction}}}
\label{sec:intro}
\begin{figure*}
	\centering
		\includegraphics[height=5.in,width=6.875in]{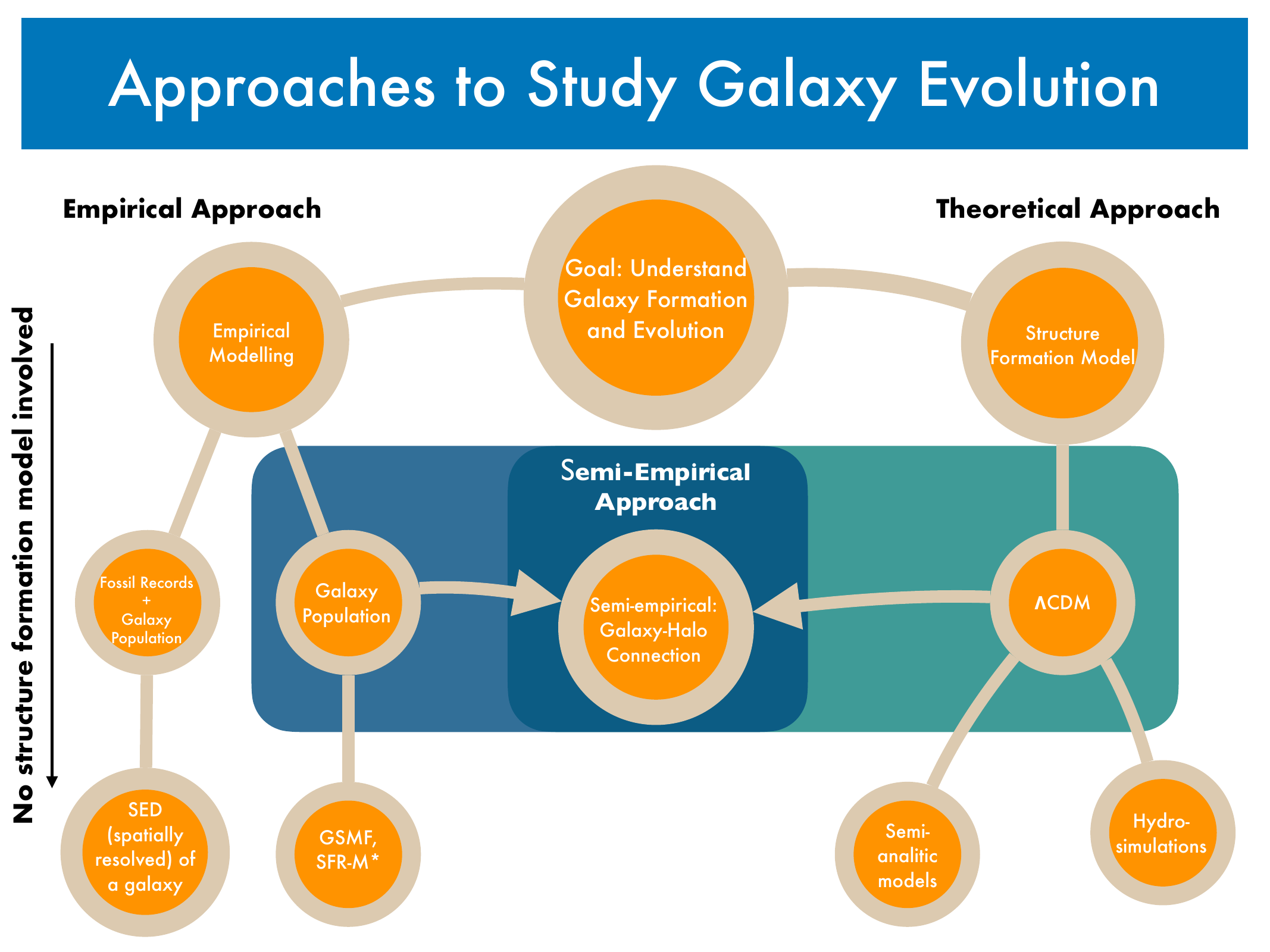}
		\caption{ Approaches to study Galaxy Evolution. {\bf The Empirical Approach} can be divided into two methods: the galaxy-by-galaxy and the population (or statistical) method. One example of the galaxy-by-galaxy method employs the fossil record approach, which utilizes the observed spectral energy distribution of galaxies to constrain properties like stellar ages. On the other hand, the population method infers the statistical evolution of galaxies by analyzing the observed evolution of the GSMF and the SFR-stellar mass relation of SFMS. While empirical approaches are firmly grounded in observational data, they lack information about the cosmological context of galaxy evolution. {\bf The Theoretical Approach}  operates under the assumption that $\Lambda$-Cold Dark Matter halos serve as the gravitational framework for galaxy formation and evolution. These models simultaneously track the hierarchical formation of dark matter halos and the astrophysical processes, employing either phenomenological recipes (semi-analytical models) or full hydrodynamics (zoom-in or full cosmological simulations). However, exploring the parameter space, subgrid physics parameterizations, and the computational expenses associated with these approaches pose challenges in simulating a broad range of physics scales and stellar masses. {\bf The Empirical Approach} operates at the intersection of the observational data and the structure formation model. The empirical approach has been proven to be a powerful and effective tool to unraveling the SFH of galaxies.}
	\label{fig:approaches}
\end{figure*}

Understanding how galaxies formed  and the intricate mechanisms governing their formation, growth, and eventual cessation of star formation (SF) are central questions in modern studies of extragalactic astronomy and cosmology. Broadly speaking, in the study of galaxy evolution, two main approaches are commonly employed: the empirical and theoretical approaches, Figure \ref{fig:approaches}.

\textbf{{\color{\darkred} The Empirical Approach:}} It is based on reconstructing galaxy evolution from observations. This can be divided into two methods: the galaxy-by-galaxy and the population (or statistical) method. The former focuses on the evolutionary diversity of individual galaxies' star-formation histories (SFH), while the latter emphasizes inferring average quantities.

\begin{itemize}
    \item \textbf{{\color{\darkred} The galaxy-by-galaxy method:}} An example of this is the fossil record approach, (Panter et al. 2007; Conroy 2013), which uses the observed spectral energy distribution of galaxies to constrain the properties of the stellar populations, particularly the distribution of stellar ages which relates to the {\it formation history} of the galaxies (see e.g., Conroy 2013; S{\'a}nchez 2020). That is, it represents the mass of all the stars in the host galaxy at the redshift of observation that had already formed by a given time. However, interpreting the cosmological assembly history regarding the main and/or individual progenitors of the galaxy can be challenging.

    \item \textbf{{\color{\darkred} The galaxy population method:}} One of such methods links the star-forming galaxies to their progenitors at high redshifts by following the stellar mass evolution implied by the star-forming main-sequence, SFMS (Leitner 2012). Another method involves examining the observed evolution of the galaxy stellar mass function (GSMF) to compare galaxies across redshifts at a fixed cumulative number density (van Dokkum et al. 2010). A third approach is a hybrid method that integrates aspects of the previous two by separately considering the populations of SFMS and quiescent galaxies  (Mu{\~n}oz \& Peeples 2015; Leja et al. 2015). However, accounting for galaxy mergers poses a challenge in the latter method, often requering the use of indirect techniques such as empirical recipes of merger rates based on projected pairs (Man et al. 2016) to estimate mass increases due to mergers.
\end{itemize}

Empirical approaches are grounded in observational data, ensuring self-consistency with observations but lacking specific reliance on any particular structure formation mode. Consequently, these models often lack information regarding the cosmological context of galaxy evolution.

\textbf{{\color{\darkred} The Theoretical Approach:}} According to the current paradigm, $\Lambda$-Cold Dark Matter halos provide the gravitational framework for galaxy formation, influenced by both hierarchical structure formation driven by gravity and the local (gas)astrophysical processes such as gas cooling, stellar-feedback driven winds, and feedback from supermassive black holes (SMBHs) and supernovae. This interdependence suggests that the growth and assembly of halos can significantly impact galaxy properties. One example of these are the semi-analytic models that utilize phenomenological recipes to capture the essential physics for galaxy formation and evolution  that are often traced through the merger trees from $N$-body dark matter simulations. Another theoretical approach involves hydrodynamical simulations, which can be either zoom-in or full cosmological simulations. These simulations simultaneously model gravity and hydrodynamics up to a certain resolution scale, requiring subgrid physics parameterizations beyond that scale, see Somerville \& Dav{\'e} (2015) for a review. 

Theoretical approaches often involve tuning a large number of parameters, which makes exploring parameter space challenging and sometimes leads to reliance on assumptions that may not accurately reflect reality. This is particularly evident in semi-analytic models.  In the case of hydrodynamical simulations, subgrid physics parameterizations introduce an additional degree of freedom that can be difficult to constrain. Moreover, their computational expense escalates when sampling a broad range of physics scales and stellar masses, presenting further challenges, see Crain \& van de Voort (2023) for a more in-depth discussion.

A third approach that has been gaining popularity over the last decade is the semi-empirical approach of the galaxy-halo connection, which combines the strengths of both empirical and theoretical approaches. 

\textbf{{\color{\darkred} The Semi-Empirical Approach:}} By operating at the intersection of observational data and the cosmological structure formation model, the semi-empirical approach (by construction) remains consistent with real-world observations and cosmological dark matter simulations. Utilizing semi-empirical modeling, researchers employ observational data on galaxy demographics, including metrics such as the GSMF and star formation rates (SFRs), with dark matter assembly histories derived from $N$-body simulations. This approach is a powerful tool that provides insights into the main mechanisms driving galaxy evolution over time by exploiting the intricate relationship between halos and galaxies. Key studies by Conroy \& Wechsler (2009); Firmani \& Avila-Reese (2010) and Behroozi et al. (2013a) have demonstrated the effectiveness of this approach in unraveling the star formation histories of galaxies. 

It is important to emphasize that the approach employed by the previous authors provides a probabilistic description of galaxy evolution, primarily driven by mass (see e.g., Rodr{\'\i}guez-Puebla et al. 2017). In other words, the galaxy assembly histories discussed in those papers represent average histories. From this perspective, it can be argued that the results presented in those papers depict the typical assembly histories of galaxies as a function of their stellar or virial masses.

In recent years, a new generation of semi-empirical modeling has emerged, utilizing the full diversity of dark matter halo assembly histories to derive case-by-case galaxies' SFHs, resulting in realistic bimodal SFR distributions. These advanced semi-empirical models consider SFHs as dependent not only on galaxy/halo mass but also on other properties. One notable example is \textsc{UNIVERSEMACHINE} by Behroozi et al. (2019; see also SHARC by Rodr{\'\i}guez-Puebla et al. 2016 and \textsc{EMERGE} by Moster et al. 2018). 

Note, however, that even in these sophisticated new models, galaxies are still represented as points (but see e.g., Rodr{\'\i}guez-Puebla et al., 2017; Zanisi et al. 2021), and only two galaxy properties are evaluated: mass and SFR. A significant next step in the semi-empirical modeling of the galaxy-halo connection will be to include other galaxy properties such as the structural properties of galaxies (e.g., Rodr{\'\i}guez-Puebla et al. 2017), metallicities (e.g., Rodr{\'\i}guez-Puebla et al. 2016) the cold gas content of galaxies (e.g., Popping et al. 2015; Calette et al. 2021) and supermassive black holes (SMBH, see e.g., Aversa et al. 2015; Zhang et al. 2023) in a self-consistent manner, enabling a deeper understanding of galaxy formation and evolution. 

In this extended contribution of the original published proceeding for the XVII Latin American Regional IAU Meeting, we introduce a new sEM-emPIRical modEl, \empire, for the galaxy-halo connection (Rodr{\'\i}guez et al., in prep.). \empire{} posits that the galaxy population follows a multivariate conditional distribution $\mathcal{P}(\vec{g}|\vec{h})$, where galaxy properties are represented by the vector $\vec{g}$ and halo properties by $\vec{h}$, which describes the galaxy-halo connection in general terms. The model self-consistently captures observed distributions of SFRs and stellar masses across time, the decomposition of the SFRs into the unobscured (far-ultraviolet) and obscured contributions (far-infrared), as well as their half-light/mass radius and central SMBHs. In this paper, we present preliminary results of the empirical SFH and star formation efficiency, SFE, of galaxies as a function of their dark matter halos and stellar masses obtained from \empire{}.

\section{ \textbf{{\color{\darkblue} The Galaxy-Halo connection}}}

We start our discussion by defining the GSMF as the number of galaxies per comoving volume with stellar mass at $M_\ast \pm dM_{\ast}/2$, denoted by $\phi_\ast(M_\ast)$. The GSMF stands as one of the most fundamental observables for studying the statistical and demographic properties of galaxies across time. All the physical mechanisms underlying the formation and evolutionary processes of galaxies are imprinted in the GSMF. Similarly, for $\Lambda$CDM, the halo mass and velocity functions serve as fundamental cosmological tools for studying the statistical properties of dark matter halos.

The simplest approach to derive the galaxy-halo connection can be achieved through the so-called (sub)halo abundance matching (SHAM). SHAM operates under the assumption of a one-to-one monotonic relationship between the stellar mass of galaxies and the halo (and/or maximum circular velocity) of dark matter halos, established by matching their corresponding number densities (Frenk et al. 1988).\footnote{This paper is perhaps one of the earliest to introduce SHAM in its current form and to empirically study the galaxy-halo connection; see their Figure 12.} In other words, for a given stellar mass, the cumulative GSMF is matched to the cumulative halo plus subhalo mass function to derive the SHMR, see Appendix \ref{secc:derivation_of_SHAM}. The generic shape of the SHMR is primarily governed by the Schechter-like functions of the galaxy stellar mass function and the power-law regime of the halo plus subhalo mass function. Other approaches, such as the halo occupation distribution (HOD, Berlind \& Weinberg 2002) and conditional stellar mass function (CSMF, Yang et al.2003), which we do not discuss here, are closely related to SHAM (see Section 2.1.2 of Rodr{\'\i}guez-Puebla et al. 2013 for further discussion). These approaches constrain the distribution of central and satellite galaxies by using the observed auto two-point correlation function (2PCF) and galaxy number densities. Nonetheless, despite the different nature between these models, they are very consistent with each other in obtaining the SHMRs at least for the local universe $z\sim0.1$ (see e.g., Rodr{\'\i}guez-Puebla et al. 2013; Behroozi et al. 2019).

A closely related quantity to the SHMR is the stellar-to-halo mass ratio, SHM ratio. The SHM ratio as a function of halo mass is well approximated by a double power-law function with a maximum at $M_{\rm vir} \sim 10^{12} M_{\odot}$, similar to the halo masses where virial shocks are expected to occur (Dekel \& Birnboim 2006). The shape of the SHM ratio is consistent with what would be expected from the processes that regulate the SFE in galaxies. In low-mass halos, SFE can be affected by energy feedback from supernova explosions and stellar winds, while AGN feedback may be important in more massive halos, causing further suppression of star formation, see Somerville \& Dav{\'e} (2015).

\subsection{Satellite galaxy-Subhalo connection}

Subhalos can lose a significant fraction of their mass due to tidal stripping. Since tidal stripping affects the subhalo more than the stars of the galaxy inside it, the stellar mass of a satellite galaxy does not trivially correlate with the current mass or velocity of its host subhalo. Therefore, when connecting galaxies to dark matter (sub)halos, as in SHAM, Reddick et al. (2013) demonstrated that the maximum halo mass or velocity ($M_{\rm peak}$ or $V_{\rm peak}$) throughout the entire history of a subhalo correlates better with the luminosity/stellar mass of the satellite galaxy it hosts. This is because assuming identical SHMRs for central and satellite galaxies, as is assumed in SHAM, leads to better reproduction of the observed 2PCF. The success of the auto-2PCF is related to the fact that the galaxy-halo connection reproduces the observed CSMFs, which describe the probability of finding a galaxy with mass $M_{\ast}$ at a given halo mass (Reddick et al. 2013). Later, Dragomir et al. (2018) showed that utilizing subhalo $V_{\rm peak}$ property accurately explains the dependence of the GSMF on environment.\footnote{Notice that if the galaxy-halo connection is independent of the environment, the modulation of the GSMF with environment is totally given by the dependence of the halo mass function on environment (Dragomir et al. 2018).} 

Alternatively, Rodr{\'\i}guez-Puebla et al. (2012, 2013) showed that assuming different SHMRs between centrals and satellite galaxies is more consistent with the observed 2PCFs and CSMFs. The authors found that when using the halo mass at the time of accretion, satellite galaxies tend to have a larger stellar mass than central galaxies for a given (sub)halo mass implying that satellites evolved similarly to central galaxies for few Gyrs.

Another alternative treatment for satellite galaxies is the inclusion of orphan galaxies. As mentioned earlier, subhalos are influenced by the tidal effects of their host halos, and whether a satellite survives depends on factors such as mass resolution and halo finders (for a discussion see Appendix B of Behroozi et al. 2019). Orphan galaxies are those whose host subhalos have lost most of their mass and are no longer resolved in the simulations. The extent to which the fraction of orphan galaxies is needed depends greatly on the resolution of the simulation and the implementation for tracking orphans. While orphans offer an intriguing possibility for accurately reproducing the clustering of galaxies, there is still ongoing discussion regarding the observational constraints and how to implement them (see e.g., Kumar et al. 2024).

\subsection{The SHMR by SF activity and the correlation between halo assembly history and SFR}
\label{secc:SHMR_by_SF}

The SHMR exhibits a tight scatter, with empirical constraints suggesting it is on the order of $\sim0.15$ dex (see, for example, Figure 2 of Porras-Valverde, 2023), yet its physical origin remains unknown. Observationally, several authors have shown that quiescent/red central galaxies tend to reside in more massive halos than star-forming/blue central galaxies at fixed stellar mass based on the kinematics of satellites (More et al., 2011), galaxy groups and clustering (Rodríguez-Puebla et al., 2015) and weak-lensing (Mandelbaum et al., 2016). This difference in halo mass is such that quiescent galaxies reside in halos that are a factor of $\sim2-3$ more massive than the host halos of star-forming galaxies in massive central galaxies, $M_\ast\gtrsim10^{11} M_{\odot}$. While this trend is robust, it remains a matter of debate whether the SHMR exhibits similar segregation.\footnote{Recall that the SHMR defines the stellar mass as a function of halo mass. The observed segregation by color/SF, however, is for the halo mass as a function of stellar mass. Inverting one relation does not equate to the other, primarily due to the effect of Eddington bias, see Equation 40 from Rodr{\'\i}guez-Puebla et al. (2015)} For instance, Moster et al. (2018, 2020) found that quiescent galaxies have higher stellar masses than star-forming galaxies at fixed halo mass. The authors proposed that the combined effect of Eddington bias and a higher fraction of star-forming galaxies at lower masses results in average halo masses being larger for quiescent galaxies than for star-forming ones at fixed stellar mass, consistent with observations. This phenomenon has been described as the inversion problem (Cui et al., 2023). Additionally, other authors have found little evidence for such segregation in the SHMR (see, e.g., Behroozi et al., 2019).

The tension among authors regarding the segregation of the SHMR is closely related to the question of the extent to which other halo properties, beyond mass, influence the star formation history of galaxies. Dark matter halo clustering not only correlate with mass but also with their assembly history, a phenomenon known as halo assembly bias (Wechsler et al. 2006). At the same time, it is well established that red/quiescent galaxies cluster more strongly than blue/star-forming galaxies (see, for example, Berti et al. 2021).

Efforts to incorporate the clustering effects from a halo's assembly history into semi-empirical models of the galaxy-halo connection have been undertaken to explore how other halo properties influence galaxies' SFRs (e.g., Hearin \& Watson, 2013). If differences in clustering between star-forming and quiescent galaxies are influenced by other halo properties, then the segregation in the SHMR may not be required, as assembly bias would boost the clustering of quiescent galaxies. Conversely, if the segregation is present, then the enhanced clustering of quiescent galaxies is due to their occupancy in more massive halos. It is worth noting that most previous studies do not separately analyze central and satellite galaxies. Including satellites adds a higher degree of complexity to the problem, as the effects of assembly bias have primarily been studied for distinct halos. Additionally, the stronger clustering effect of quiescent galaxies is mainly attributed to satellites (see Kakos et al., 2024, for further discussion)

In Kakos et al. (2024), we investigated the aforementioned tension by analyzing the clustering of central galaxies with $M_\ast > 10^{10} M_{\odot}$ as a function of their distance from the SFMS, while considering bins of stellar masses. We compared these results to models for the galaxy-halo connection. In our study, we developed two models where the sSFR depends on halo assembly history (utilizing halo accretion and concentration), and one model consistent with the segregation in the SHMR. One notable finding was that the observed auto-2PCFs showed little dependence on sSFR, and all models replicated this trend. This suggests that the impact of other halo properties may not be easily discernible from the auto-2PCFs. In contrast, cross-correlation of centrals with satellites exhibited signatures that only the segregation model was able to reproduce. This implies that there is little net correlation between halo assembly history and sSFR of central galaxies. 

The discussions presented in this section provide important contextual considerations that are essential for introducing our model in the subsequent section.

\begin{figure*}
	\centering
		\includegraphics[height=6.5in,width=6.5in]{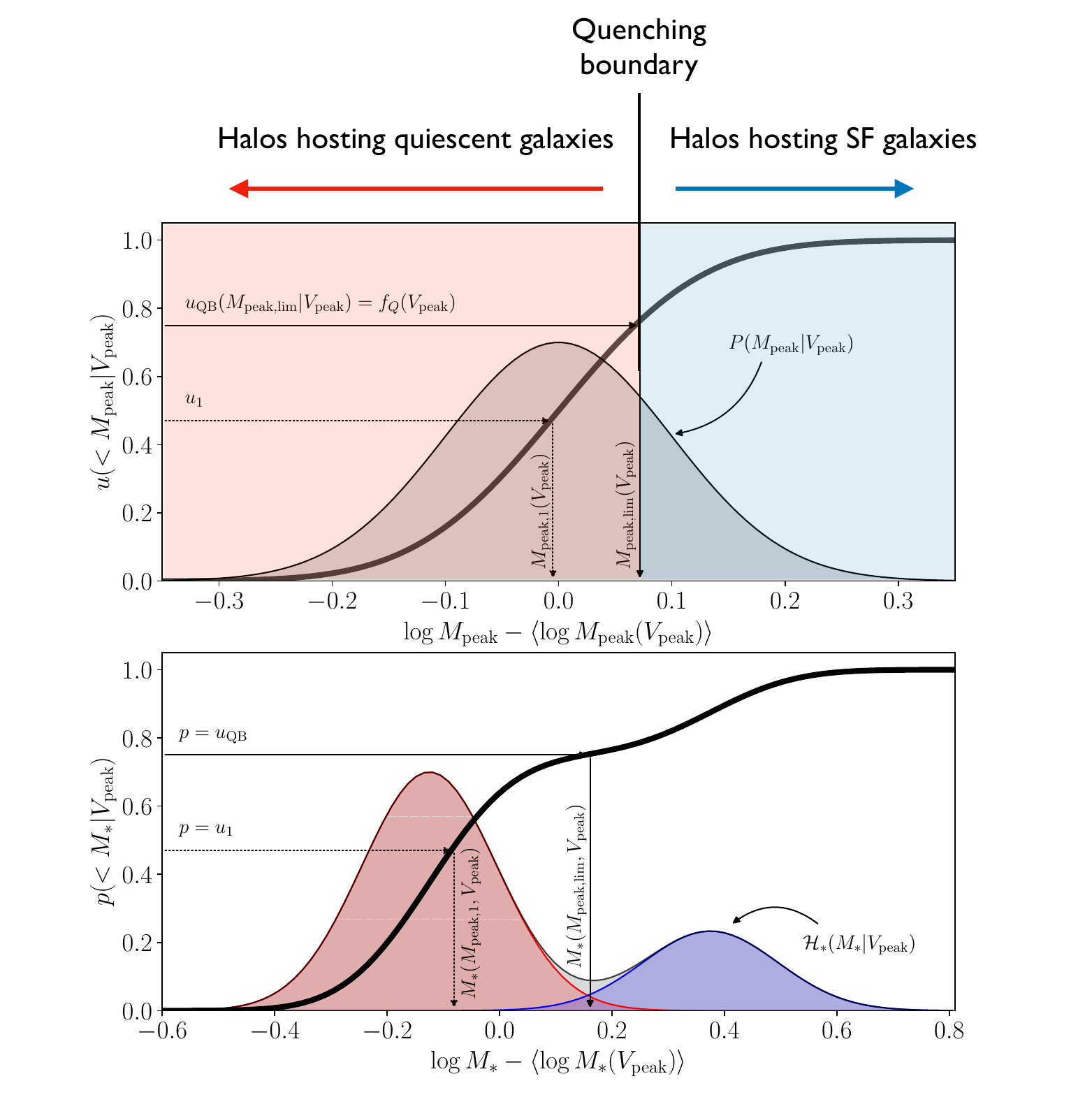}
		\caption{ The relationship between $M_\ast$-$V_{\rm peak}$-$M_{\rm peak}$. {\bf Upper Panel:} The conditional cumulative probability $u(<M_{\rm peak}|V_{\rm peak})$ for (sub)halos with masses lower than $M_{\rm peak}$ and (sub)halos with a peak velocity $\log V_{\rm peak}\pm d\log V_{\rm peak}/2$ is shown with the thick solid line. The thin solid line shows the conditional cumulative probability distribution $\mathcal{P}_{V}(M_{\rm peak}|V_{\rm peak})$ indicating that a (sub)halo with a peak velocity $\log V_{\rm peak}\pm d\log V_{\rm peak}/2$ has a value of $M_{\rm peak}$ within the range of $\log M_{\rm peak}\pm d\log M_{\rm peak}/2$. We present two values for $u = u_1$ and $u_{\rm QB}$, the latter one corresponds to the halo peak mass $M_{\rm peak, lim}$ which defines a quenching boundary, see the text for details. {\bf Bottom Panel: } The conditional cumulative probability $p(<M_{\ast}|V_{\rm peak})$ for galaxies with masses lower than $M_{\ast}$ and (sub)halos with a peak velocity $\log V_{\rm peak}\pm d\log V_{\rm peak}/2$, represented here as the thick solid line. The thin solid line shows the the bimodal conditional probability distribution $\mathcal{H}_\ast( M_{\ast} | V_{\rm peak})$ with one mode representing star-forming galaxies and a second mode quiescent galaxies. The corresponding values of $p = u_1$ and $p = u_{\rm QB}$ and their corresponding stellar masses are shown with the dotted and solid arrows respectively. The figure illustrates how \empire{} assigns stellar masses to dark matter (sub)halos. At a given $V_{\rm peak}$ \empire{} assumes that star-forming galaxies have larger $M_{\rm peak}$ and $M_\ast$ than quiescent galaxies following the recent results by Kakos et al. (2024).}
	\label{fig:empire_rules}
\end{figure*}

\section{ \textbf{{\color{\darkblue} The semi-empirical modeling for the galaxy-halo connection: \empire}}}

\empire{} is a semi-empirical model for the galaxy-halo connection that models the SHMR of individual galaxies based on the growth of dark matter (sub)halos obtained from N-body simulations. It tracks the assembly histories of both star-forming and quiescent galaxies, resulting in a bimodal distribution of SFRs among galaxies. This is achieved by using $V_{\rm peak}$ to establish a connection between a galaxy's stellar mass and its SFH, while $M_{\rm peak}$ is linked to the likelihood of a galaxy undergoing quenching based on its host dark matter halo's assembly history. For further details, refer to Rodr\'iguez-Puebla et al. (in prep.). Next, we briefly describe some important aspects from \empire{}. It is important to note that the way we are correlating star-forming and quiescent galaxies to their host dark matter halos is strongly influenced by recent findings from Kakos et al. (2024) with the goal to find realistic SHMR and the observed spatial clustering of galaxies, see Section \ref{secc:SHMR_by_SF}.

In broad terms, \empire{} operates under the assumption that at a given $V_{\rm peak}$, halos hosting star-forming galaxies tend to have greater stellar mass than those hosting quiescent galaxies. Conversely, halo mass acts as an indicator of quiescence in the galaxy. To assign star-forming and quiescent galaxies to their respective host halos, \empire{} utilizes the $M_{\rm peak}$ and $V_{\rm peak}$ values tabulated in halo catalogs from $N$-body simulations. At each snapshot, characterized by its redshift $z$, and for every (sub)halo, \empire{} computes the following cumulative probability:
	\begin{equation}
		u = \int_{0}^{M_{\rm peak}} \mathcal{P}_{V}( M'_{\rm peak}  | V_{\rm peak}) d\log M'_{\rm peak}.
		\label{eq:random_sfid}
	\end{equation}
Here $\mathcal{P}_{V}(M_{\rm peak}|V_{\rm peak})$ represents the conditional probability distribution function, indicating that a (sub)halo with a peak velocity $\log V_{\rm peak}\pm d\log V_{\rm peak}/2$ has a value of $M_{\rm peak}$ within the range of $\log M_{\rm peak}\pm d\log M_{\rm peak}/2$. It is noteworthy that if the fraction of halos hosting quiescent galaxies is denoted as $f_Q(V_{\rm peak})$, then:
\begin{equation}
	u(M_{\rm peak, lim},V_{\rm peak}) = f_{\rm Q}(V_{\rm peak}).
	\label{eq:quenching_boundary}
\end{equation}
(Sub)halos above the $M_{\rm peak, lim}$ threshold will host star-forming galaxies. In other words, $M_{\rm peak, lim}(V_{\rm peak})$ defines a \emph{quenching boundary} in the $M_{\rm peak}$ and $V_{\rm peak}$ plane for the galaxies in \empire{}, as illustrated in the upper panel of Figure \ref{fig:empire_rules}

The next step is to define $p(M_{\ast}|V_{\rm peak})$ as the cumulative probability that a galaxy will have a stellar mass lower than $M_{\ast}$ given its $V_{\rm peak}$:
\begin{equation}
	 p(M_{\ast}|V_{\rm peak}) =  \int_{0}^{M_{\ast}} \mathcal{H}_\ast( M'_{\ast}  | V_{\rm peak}) d\log M'_{\ast}.
	\label{eq:percentile_j}
\end{equation}
Here, $\mathcal{H}_\ast( M_{\ast} | V_{\rm peak})$ is a bimodal conditional distribution, with one mode representing star-forming galaxies and the other mode representing quiescent galaxies (see the bottom panel of Figure \ref{fig:empire_rules}). This conditional distribution is the galaxy-halo connection. As illustrated in the figure, for a given $V_{\rm peak}$, star-forming galaxies generally have a larger stellar mass compared to quiescent galaxies. Stellar masses are determined by solving the following equation for $M_{\ast}$:
\begin{equation}
	p(M_\ast|V_{\rm peak}) = u(M_{\rm peak}|V_{\rm peak}).
	\label{eq:prob_ms}
\end{equation} 
Thus, stellar masses result a function dependent on $V_{\rm peak}$ and $M_{\rm peak}$:
\begin{equation}
    M_{\ast} = M_{\ast}(M_{\rm peak},V_{\rm peak}),
\end{equation}
that is, in \empire{}, the stellar masses of galaxies depend on two halo properties. The entire process described above is illustrated in Figure \ref{fig:empire_rules}. Notice that while not explicitly shown, this process is conducted at each redshift of the simulation. Additionally, it is worth mentioning that the described process is executed separately for central and satellite galaxies. Specifically, the fraction of quiescent satellites exceeds that of central galaxies at lower halo masses, while at higher masses, both fractions are equal. At higher redshifts, $z \sim 3$, the fraction of quiescent satellite galaxies is nearly identical to that of centrals. Additionally, we assume that centrals and satellites follow different mean $M_\ast$-$V_{\rm peak}$ relationships.

\begin{figure*}
	\centering
		\includegraphics[height=4.7in,width=6.5in]{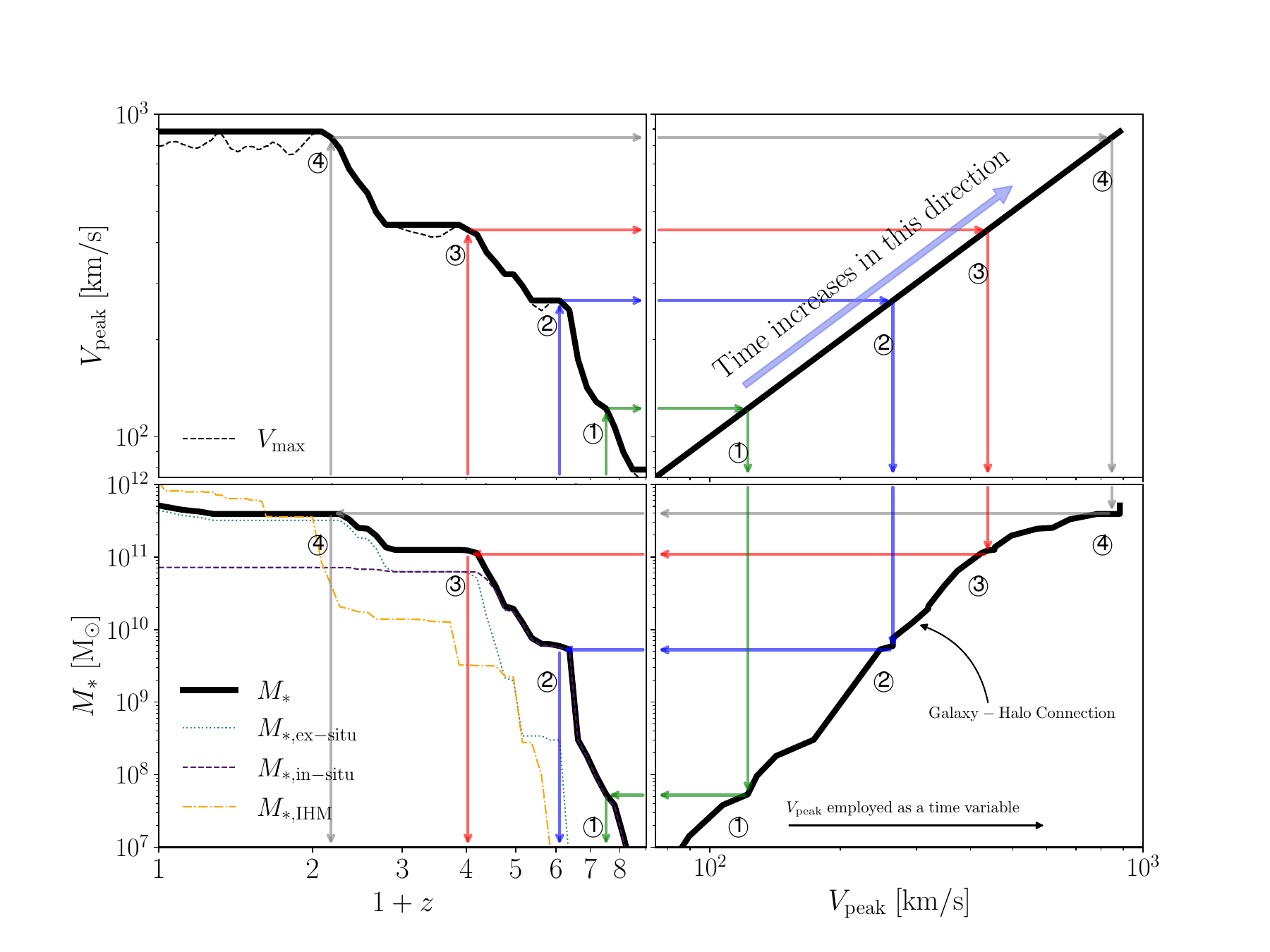}
			\caption{ The process of inferring the mass assembly history of a massive galaxy within a dark matter halo with a mass of $M_{\rm peak} = 2 \times 10^{14} M_{\odot}$. We highlight four key points in the history of the dark matter halo/galaxy to illustrate their correspondence to the galaxy's assembly history. In the clockwise direction: {\bf Upper Left Panel: }  This panel depicts the evolution of the main progenitor of $V_{\rm peak}$ (solid black line) and $V_{\rm max}$ (dashed line). {\bf Upper Right Panel: } Here, we note that $V_{\rm peak}$ has a more monotonic behavior over time compared to $V_{\rm max}$. This behavior underscores the suitability of $V_{\rm peak}$ as a proxy for stellar masses, given its correlation with time.  {\bf Bottom Right Panel: } The galaxy-halo connection.  In semi-empirical modeling, galaxies evolve based on $V_{\rm peak}$ according to the rules established in Figure \ref{fig:empire_rules} and halos serve as ``cosmological clocks" that will establish the pace at which galaxies will grow. {\bf Bottom Left Panel: } The resulting evolution of the mass assembly of the main progenitor of the galaxy, represented by the solid black line. This evolution is decomposed into its in-situ (dashed line), ex-situ (dotted line), and intra-halo mass (dot-dashed line) components.
 		}
	\label{fig:examples_of_ms_history}
\end{figure*}

\begin{figure*}
	\centering
		\includegraphics[height=4.36in,width=6in]{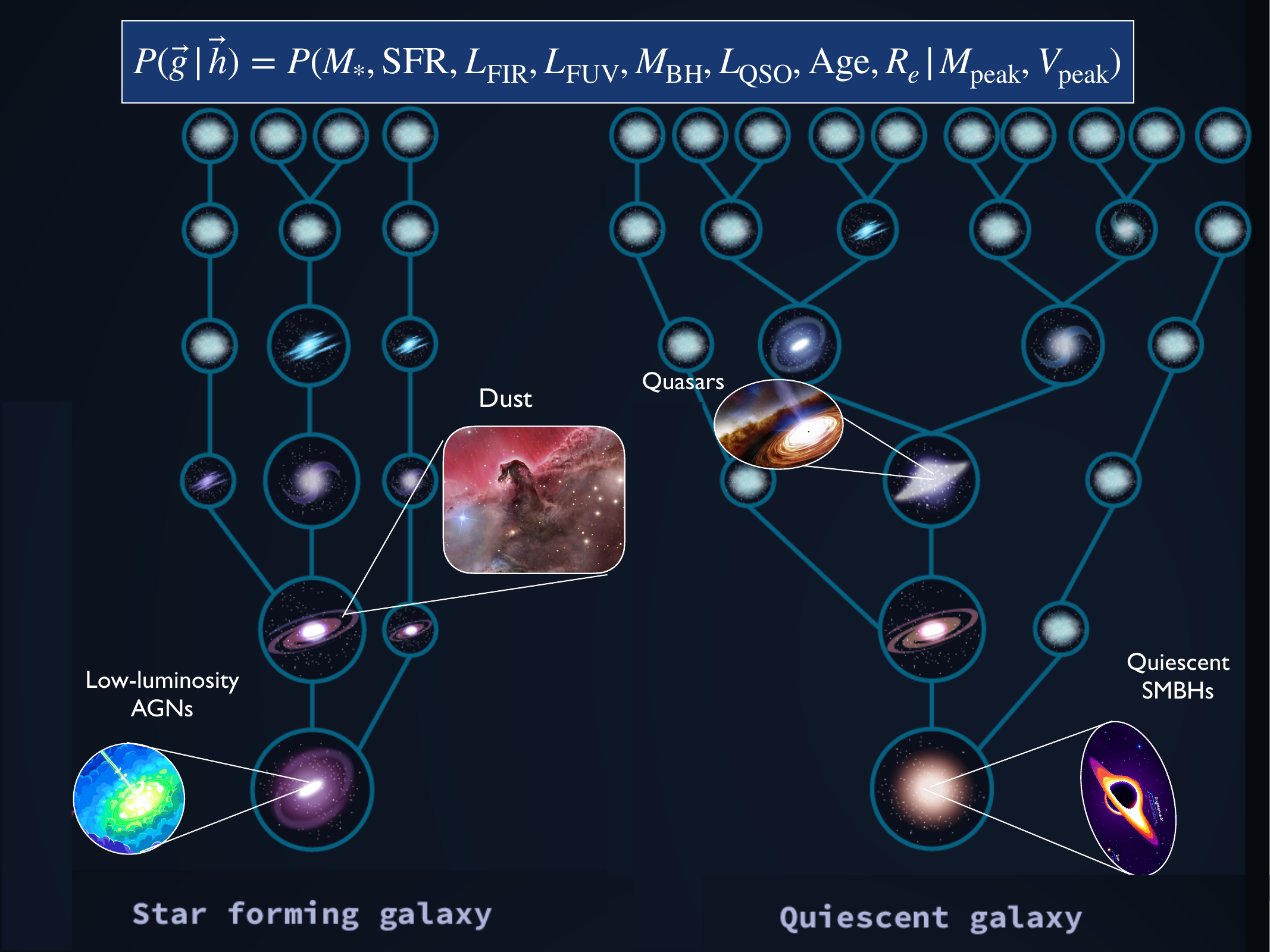}
		\caption{This schematic figure illustrates the assembly history of both star-forming and quiescent galaxies. In the context of \empire{}, this model characterizes the multivariate distribution $\mathcal{P}(\vec{g}|\vec{h})$, which defines the relationship between halos (represented by $\vec{h}$) and galaxies (represented by $\vec{g}$). This distribution encapsulates the various properties of galaxies as influenced by their underlying halos, providing insight into the galaxy-halo connection.}
	\label{fig:cartoon}
\end{figure*}

\begin{figure*}
	\centering
		\includegraphics[height=1.95in,width=3.in]{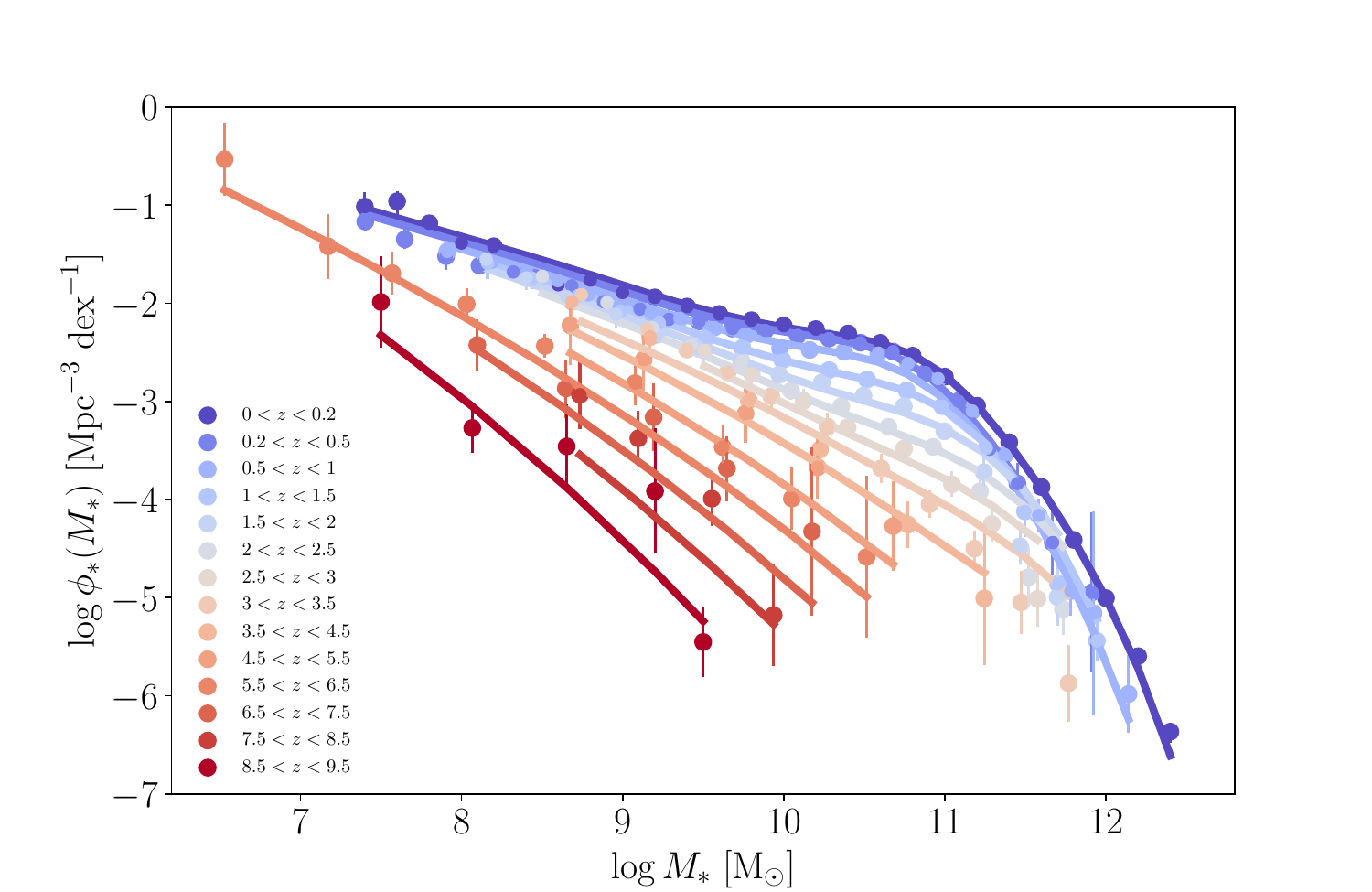}
  		\includegraphics[height=1.95in,width=3.in]{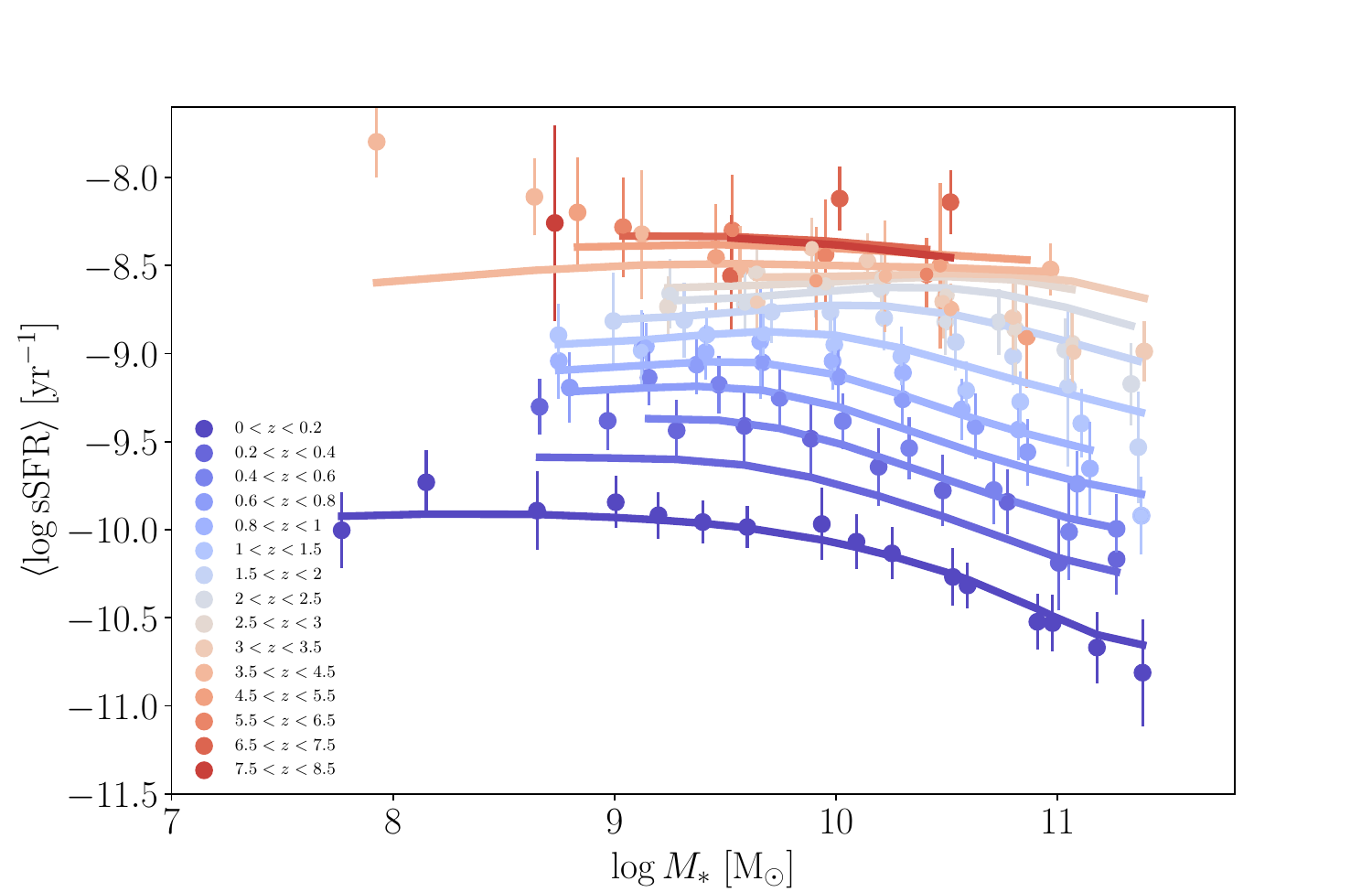}
			\caption{
			Observed data (shown as circles with error bars) and the corresponding best-fitting model generated by \empire{} (represented by solid lines). In the left panel, the plot illustrates the GSMF for all galaxies within the redshift range of $0 < z < 9$. The right panel focuses on the specific star formation rate as a function of stellar mass for star-forming main sequence galaxies within the redshift range of $0 < z < 8.5$.
 		}
	\label{fig:constrains}
\end{figure*}

Figure \ref{fig:examples_of_ms_history} illustrates the process of inferring the mass assembly history of a massive galaxy in a dark matter halo of mass of $M_{\rm peak} = 2 \times 10^{14} M_{\odot}$. This figure should be read clockwise starting from the upper left panel. In the upper left panel, the evolution of $V_{\rm peak}$ is shown with the black solid line, while the dashed line depicts the evolution of $V_{\rm max}$. Notice that $V_{\rm peak}$ serves as a better proxy for time than $V_{\rm max}$ since it exhibits a monotonic behavior, as shown in the upper left panel. This characteristic is desirable because stellar mass is an increasing monotonic function that correlates with time. Therefore, it is not only reasonable but also logical to conclude that $V_{\rm peak}$ should correlate better with $M_\ast$ than $V_{\rm max}$ (see also Reddick et al. 2013). 

The lower right panel illustrates the evolution of a galaxy around the $M_\ast-V_{\rm peak}$ obtained when solving Equation (\ref{eq:prob_ms}) redshift-by-redshift. For illustrative purposes, we have identified four points in the history of the dark matter halo to show how they correspond to the assembly history of the galaxy. These points are indicated by arrows connecting the different panels. In the lower left panel, the resulting assembly stellar mass history (solid black line) is displayed when converting $V_{\rm peak}$ into a time variable, i.e., $z$. This entire process lies at the core of semi-empirical modelling, where the galaxy-halo connection, bottom right panel of Figure \ref{fig:examples_of_ms_history}, is utilized to infer the assembly history, bottom left panel, of galaxies by leveraging the assembly history of dark matter halos as ``cosmological clocks" that will establish the pace at which galaxies will grow, upper right panel. 

\empire{} adopts an approach similar to that described in Rodr{\'\i}guez-Puebla et al. (2017, see also Conroy \& Wechsler 2009) to derive the mass assembly and SFH of galaxies. Specifically, this method involves using the trajectory described by $M_\ast$ to calculate the differences between two snapshots. From this, we can estimate the amount of stellar mass formed at any interval of time, which is expected to be proportional to the SFH of the galaxy. 

In general, \empire{} assumes that galaxies assemble their masses through two channels:
\begin{itemize}
	\item \textbf{{\color{\darkred} \emph{in-situ}}} star formation: This results from the birth of new stars due to the collapse of molecular clouds in the galaxies or disk instabilities.	
	
	\item \textbf{{\color{\darkred} \emph{ex-situ}}} processes: This results from galaxy mergers, the infall of stars from the intra-halo medium, and/or stars stripped from surviving galaxies.
\end{itemize}
Details will be provided in Rodr{\'\i}guez-Puebla et al. (in prep.). Notice we display the in-stiu and ex-situ components for the galaxy shown in Figure \ref{fig:examples_of_ms_history}. This galaxy becomes quiescent around $z\sim3$ when the in-situ stellar mass growth ceases (dashed line in the bottom panel) and the ex-situ growth takes over (dotted line). Additionally, we display the growth of the intra-halo mass (IHM, dot-dashed line).

Figure \ref{fig:constrains} shows the best fitting models of \empire\ to the observed GSMF from $z\sim0$ to $z\sim9$ and the sSFR for SFMS galaxies as a function of $M_\ast$ from $z\sim0$ to $z\sim8$. These observables are two of the most relevant constraints used to derive the SFH of galaxies, which will be discussed in the next section. Another relevant observable is the fraction of quiescent galaxies as a function of $M_\ast$ and $z$, $f_{\rm Q}(M_\ast,z)$, which allows constrains in Eq. \ref{eq:quenching_boundary}.

In general, \empire{} utilizes a comprehensive set of observational constraints to establish the connection between several galaxy properties, SMBHs, and dark matter (sub)halos. Consequently, each (sub)halo within the model encompasses 13 properties that define the galaxy-SMBH-halo connection. These properties will be discussed in more detail in Rodr{\'\i}guez-Puebla et al. (in prep.). Below, we outline the properties derived in \empire{}:
\begin{itemize}
    \item Stellar mass: $M_\ast$,
    \item The fraction of mass formed by in-situ process: $f_{\rm ins}$,
    \item The stellar mass formed from mergers: $M_{\ast, \rm mrg}$,
    \item Intra-halo mass: IHM,
    \item Star formation rate: SFR,
    \item Contribution of the dust-obscured star formation: $f_{\rm obs}$,
    \item Far ultraviolet luminosity: FUV,
    \item Far infrared luminosity: FIR,
    \item Half-mass radius: $R_\ast$,
    \item Half-light radius at 5000 \AA: $R_e$,
    \item Black hole mass: $M_{\rm BH}$,
    \item Black hole accretion rate: $\dot{M}_{\rm BH}$,
    \item Radiative efficiency: $\epsilon_{\rm acc}$,
\end{itemize}
Figure \ref{fig:cartoon} schematically illustrates how these properties are traced back along the merger trees of dark matter halos, particularly when distinguishing between star-forming and quiescent galaxies. All these properties are encapsulated by the vector $\vec{g}$, while the ordered pair vector $\vec{h}$ is defined as $\vec{h} = (V_{\rm peak},M_{\rm peak})$. Thus, the primary objective of \empire\ is to empirically determine the multivariate conditional distribution function $\mathcal{P}(\vec{g}|\vec{h})$, which characterizes the connection between halo properties (denoted by $\vec{h}$) and galaxy propreties ($\vec{g}$), as described in the Introduction.

Finally, preliminary results on the dust-obscured contribution to the SFRs have already been presented in Nava-Moreno et al. (2024), while results on the evolution of central galaxies in clusters and their SMBHs were presented in Montenegro-Taborda et al. (2023).

\section{ \textbf{{\color{\darkblue} Results: The co-evolution between galaxies and dark matter halos}}}

\label{sec:galaxy_halos}

\begin{figure*}
	\centering
	\includegraphics[height=3in,width=6in]{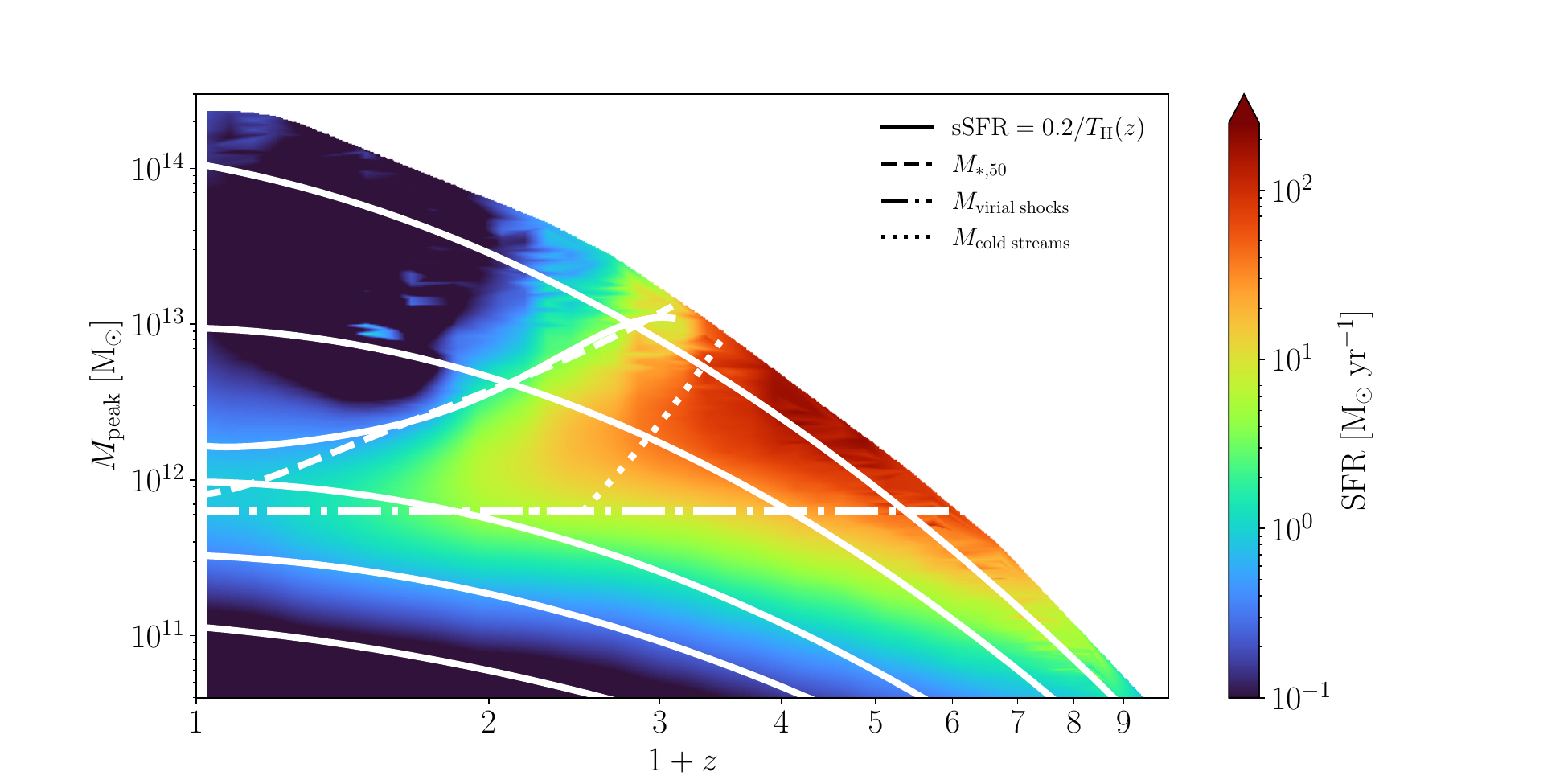}
	\includegraphics[height=3in,width=6in]{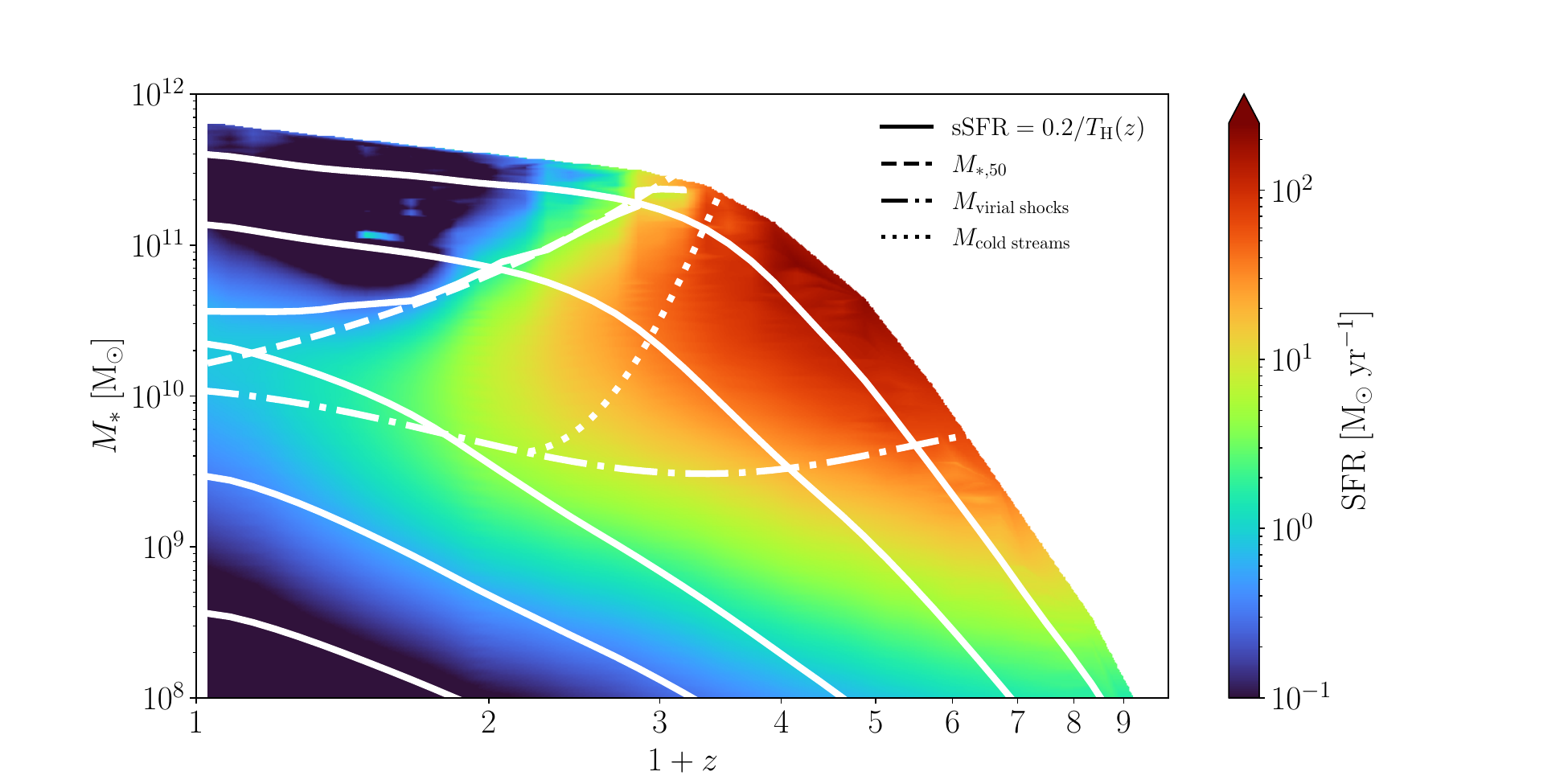}
		\caption{ Star formation histories of central galaxies as a function of $M_{\rm peak}$ (upper panel) and $M_\ast$ (bottom panel) for the progenitors of halos/galaxies at $z=0$. The progenitors of dark matter halos at $z=0$ with masses $M_{\rm peak} = 10^{11}, 10^{11.5}, 10^{12}, 10^{13}$ and $10^{14} {\rm M}_{\odot}$ are indicated by white solid lines. The solid line indicates the threshold when the sSFR$=0.2/T_{\rm H(z)}$ where $T_{\rm H}$ is the Hubble time at $z$, the dashed line indicates when 50$\%$ of galaxies are actively forming stars, and the dot-dashed and dotted lines represent the mass limits for coldstreams and shock-heated gas, respectively. The star formation histories of central galaxies in the most massive halos reached their SFR peak below the cold stream mass limit, eventually undergoing quenching. 
 	}
	\label{fig:sfh}
\end{figure*}

\begin{figure*}
	\centering
	\includegraphics[height=3in,width=6in]{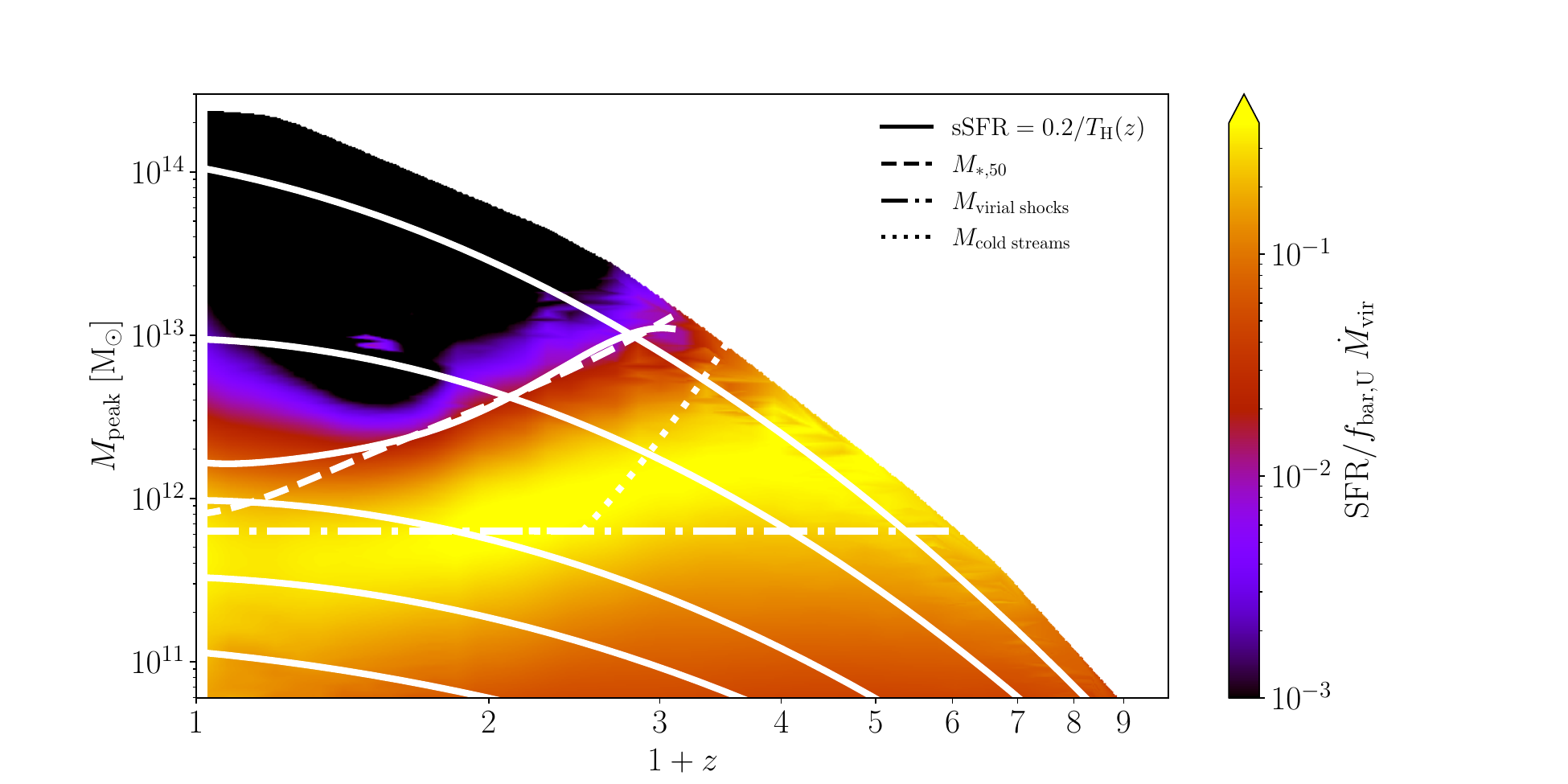}
	\includegraphics[height=3in,width=6in]{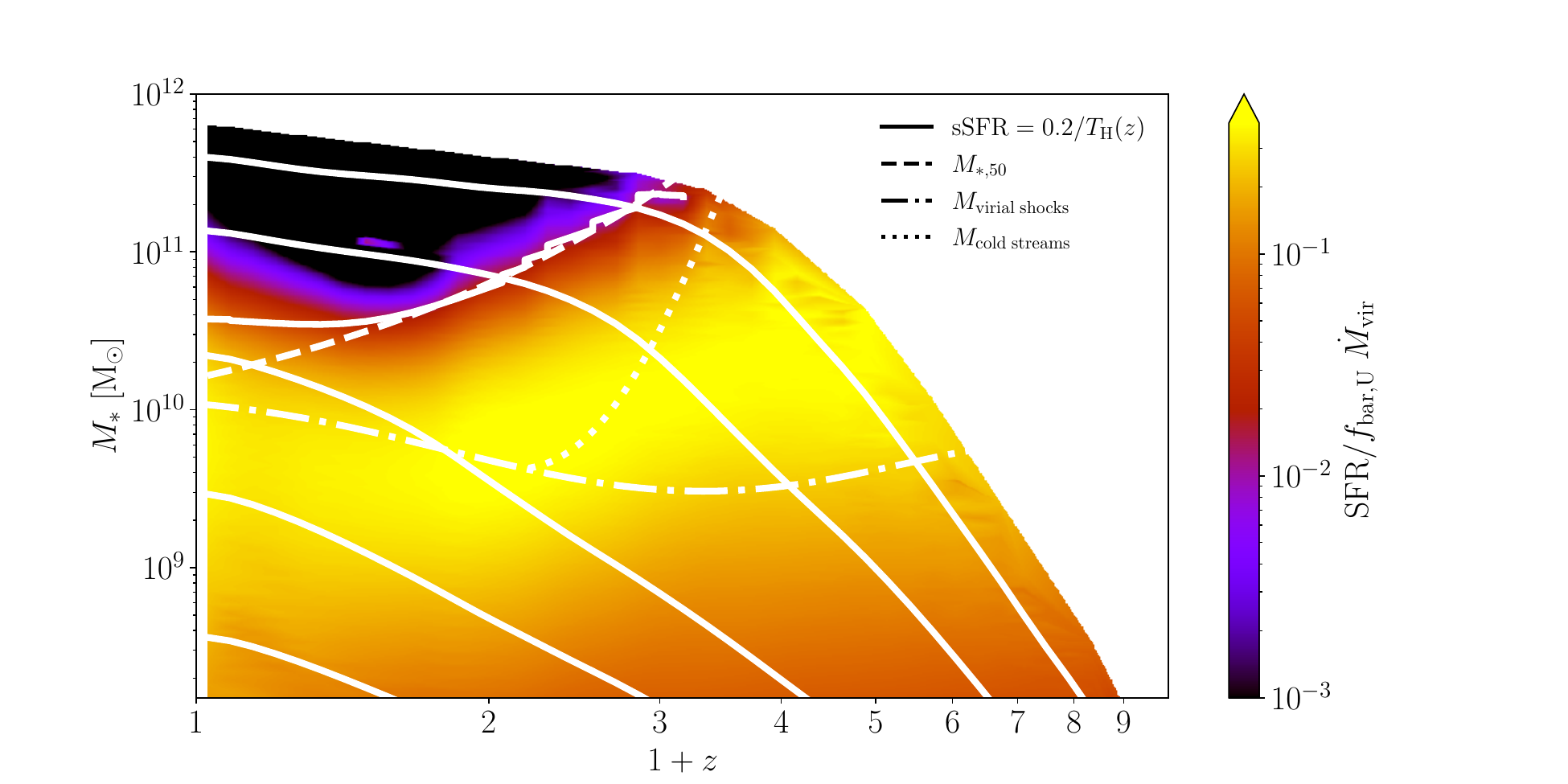}
		\caption{ Star formation efficiency of central galaxies as a function of $M_{\rm peak}$ (upper panel) and $M_\ast$ (bottom panel)  for progenitors of halos/galaxies at $z=0$. The symbols and lines are similar to those in Figure \ref{fig:sfh}. The star formation efficiency exhibits a strong dependence on halo mass and a moderate dependence on the redshift of central galaxies. The maximum efficiency is approximately a factor of $\sim1.5-2$ below $M_{\rm vir ; shocks}$ for $z\lesssim1$. Conversely, at higher redshift this maximum tends to be located towards higher masses, approximately $M_{\rm vir}\sim 2\times 10^{12} M_{\odot}$. Moreover, at higher redshifts than $z\sim2$ the peak of star formation efficiency comfortably aligns within the region characterized by cold streams.
 	}
	\label{fig:sf_efficiency}
\end{figure*}

Figure \ref{fig:sfh} presents the SFHs of galaxies ranging from dwarf galaxies, $M_*=10^{8} {\rm M}_{\odot}$, to giant ellipticals at the centers of big clusters, $M_*\sim 7 \times 10^{11} {\rm M}_{\odot}$. The thick white lines depict the average growth history of central galaxies in dark matter halos with masses at $z=0$ of $M_{\rm peak} = 10^{10.5},  10^{11}, 10^{12}, 10^{13}$ and $10^{14} {\rm M}_{\odot}$ utilizing merger trees from the SmallMultidark-Planck (SMDP) simulation (see Klypin et al. 2016 for details).

Galaxies, ranging from dwarfs to intermediate masses, $M_*=10^{8} - 10^{10} {\rm M}_{\odot}$, have been forming stars at a roughly constant rate, approximately $ 0.1$ ${\rm M}_{\odot} / {\rm yr}$ for the smallest dwarfs, $M_\ast<10^{9} M_{\odot}$, $\sim 0.5$ ${\rm M}_{\odot} / {\rm yr}$ for small galaxies, $M_\ast\sim 5\times 10^{9} M_{\odot}$ and $\sim 5$ ${\rm M}_{\odot} / {\rm yr}$  for the intermediate ones, $M_\ast\sim 10^{10} M_{\odot}$. Milky Way (MW)-sized galaxies exhibit a more complex SFH compared to dwarfs and intermediate galaxies. The SFH of MW-sized galaxies increased by a factor of $\sim20$ between $z\sim 3.5$ and $\sim1$ reaching a maximum of around $10$ ${\rm M}_{\odot} / {\rm yr}$  before declining.  In contrast, more massive galaxies, on average, have undergone elevated periods of star formation. The progenitors of giant ellipticals reached an average of  $\sim 200$ ${\rm M}_{\odot} / {\rm yr}$ between $z\sim3-4$, followed by a rapid decline. 

The formation process of massive galaxies with $M_{\ast} > 10^{11} $ $M_\odot$ ( $M_{\rm peak} \gtrsim 10^{13} $ $M_\odot$) is complex and  involves distinct stages. Initially, when their halos had masses below $M_{\rm peak} \sim 6 \times 10^{11} $ $M_\odot$  (approximately $M_{\ast}  \sim 2-4 \times 10^{9} $ $M_\odot$ for all $z$) gas accretion was highly efficient due to the short cooling time compared to the dark matter's dynamical time. During this phase, gas freely fell into the galaxy leading to an elevated period of star formation and causing galaxies to grow by approximately two orders of magnitude from $z\sim6-8$ to $z\sim3-5$. 

As halos reached masses around $M_{\rm peak} \sim 6 \times 10^{11} $ $M_\odot$ the incorporated gas was heated by virial shocks, resulting in longer cooling times. The gas temperature rose to levels where baryonic cooling efficiency decreased, limiting cold gas from forming stars. If the cooling time was shorter than the halo's dynamical time, the gas fell (within a cooling radius) and settled in the galaxy (White \&\ Frenk 1991). Alternatively, if the cooling time was longer, the gas was unable to radiate thermal energy, forming a (quasi-) hydrostatically balanced atmosphere (halo quenching, Dekel \&\ Birnboim 2006). Surprisingly, the empirical evidence that is deduced from Fig. \ref{fig:sfh} shows that galaxies continued to exhibit elevated periods of star formation even as most of the progenitors of today's massive galaxies crossed the halo mass quenching transition (dot-dashed line). 

It is worth noting that at intermediate redshifts, $z \sim2-4$, hot halos could be penetrated by cold streams (Kere{\v{s}} et al. 2005;  Dekel \&\ Birnboim 2006), supplying sufficient gas to sustain star formation in massive halos between $z\sim2-5$. This gas supply facilitated not only mass growth but also an increase in central density (see Lapiner et al. 2023 for a recent discussion), contributing to the growth of SMBHs, coinciding with the peak epoch of high SMBH accretion (Kormendy \&\ Ho 2013).\footnote{We note that EMPIRE includes the model of SMBHs that are co-evolving with galaxies and halos, and its results agree with observations by construction.} As cold streams became less efficient and halo quenching took over (the region above the dotted line in Figure \ref{fig:sfh}), the SFH experienced a strong suppression by $\sim2-3$ orders of magnitude below $z\sim1-2$. Specifically, the sSFR dropped below $0.2/T_{\rm H(z)}$, a nominal threshold for quenching, where $T_{\rm H}$ is the Hubble time at $z$, leading these galaxies to become part of the population of quiescent galaxies (solid and dashed lines). Despite the suppression of star formation, galaxies were still able to grow by a factor of $\sim2-3$ through galaxy mergers since $z\sim2$, see also Figure \ref{fig:examples_of_ms_history}.  

We now shift our analysis towards examining the star formation efficiency (SFE), which is defined as the ratio between the SFR and the baryon accretion rate, $f_{\rm bar} \dot{M}{\rm vir}$, where $f_{\rm bar} = \Omega_{\rm bar} / \Omega_{\rm matter} = 0.156$ based on the cosmological model adopted in this study. Figure \ref{fig:sf_efficiency} illustrates the evolution of the SFE for galaxies at $z=0$, presented in a similar format to Figure \ref{fig:sfh}. At a fixed redshift, the SFE shows a strong dependence on both halo and stellar mass. Particularly, quiescent galaxies residing in massive halos exhibit lower SFE values. This is expected as quiescent galaxies halt their star formation, while their host halos continue to grow hierarchically. Indeed, this interpretation aligns with the segregation in the SHMR discussed in Section \ref{secc:SHMR_by_SF}. Furthermore, the correlation becomes even stronger as a function of stellar mass, owing to the pronounced double-power law-like shape of the SHMR.

At a fixed halo mass, the SFE undergoes moderate changes with redshift, contrasting with findings from previous studies (Behroozi et al., 2013b). The peak of the SFE occurs at a halo mass slightly below $M_{\rm vir \; shocks}$, within a factor of $\sim1.5-2$, up to $z\sim 1$. However, at higher redshifts, the peak tends to shift towards higher masses, approximately $M_{\rm vir}\sim2\times10^{12} M_{\odot}$. Notably, for redshifts larger than $z \sim 2$, the peak resides well withing within the region where cold streams can supply gas, sustaining star formation and ensuring these halos remain as the most efficient ones. This is consistent with the observation that these halos host galaxies with the highest amounts of SFRs. The relationship between SFE and redshift becomes even more complex as a function of stellar mass, exhibiting significant variations. 

\section{ \textbf{{\color{\lightblue} Conclusions}}}

Semi-empirical modeling is a sophisticated tool that operates at the intersection of observational data on galaxy demographics and cosmological structure formation models, ensuring consistency with real-world observations and theory. Furthermore, the sophistication of these models has reached the level of describing the galaxy-halo connection on a halo-by-halo basis in $N$-body simulations (see e.g., Behroozi et al., 2019, and Rodr{\'\i}guez-Puebla et al., in prep). This advancement enables researchers to address fundamental questions, such as the origin of scatter around the SHMR, making them invaluable for studying galaxy formation and evolution from a cosmological standpoint. 

In this paper we exploited \empire{} a semi-empirical model for the galaxy-halo connection  (Rodr{\'\i}guez-Puebla et al. in prep) to understand the complex interplay between galaxies and their halos, shedding light on the SFH and SFE of central galaxies over cosmic time. The presented SFHs in Figure \ref{fig:sfh} and SFEs in Figure \ref{fig:sf_efficiency} unveil the varied evolutionary trajectories of galaxies, spanning from dwarfs to massive ellipticals. Particularly noteworthy are the discernible growth stages observed in the progenitors of massive galaxies. The data aligns with the notion that, at higher redshifts, cold streams played a key role in sustaining star formation in massive galaxies, while virial shock heating might become more prominent at lower redshifts.

Finally, the significance and true power of semi-empirical modeling extends beyond the development of accurate models that align with observations (i.e. having robust determinations of the stellar-to-halo mass relations, SHMR). It serves as a valuable phenomenological tool for comprehending the growth of galaxies through the projection of various galaxy observables related to their evolution to their host dark matter halos. This does not imply that semi-empirical modeling of the galaxy-halo connection should replace more physically motivated studies, as those mentioned previously. Instead, it underscores the importance of synergy between these approaches to achieve a more comprehensive and accurate understanding of galaxy formation and evolution.

\begin{appendix}

\section{ \textbf{{\color{\lightblue} A simple Derivation for SHAM}}}
\label{secc:derivation_of_SHAM}

As described in the main text, the GSMF is defined as the number of galaxies per comoving volume in $M_\ast\pm dM_{\ast}/2$. The cumulative number density of galaxies is therefore given by
\begin{equation}
    n_{\ast}(>M_{\ast}) = \int_{M_\ast}^{\infty} \phi_\ast (M'_\ast) \; dM'_\ast.
\end{equation}
 The probability of finding a galaxy in a infinitesimal volume $\delta V$ is given by (Peebles 1980)
 \begin{equation}
     \delta P_\ast = n_{\ast} \delta V.
 \end{equation}
In a similar way, for dark matter halos we can define 
\begin{equation}
   \delta P_{\rm h} = n_{\rm h} \delta V 
\end{equation}
where $n_{\rm h}$ is the cumulative number density of $\mathcal{H}$, where it can be either the cumulative halo mass function $n_{\rm M}(M_{\rm peak})$ or the maximum circular velocity one, $n_{\rm V}(V_{\rm peak})$. 

Both galaxies and halos reside within the same infinitesimal volume $\delta V$ if only if their probabilities are equal $\delta P_\ast = \delta P_{\rm h}$ therefore:
\begin{equation}
    \int_{M_\ast}^{\infty} \phi_\ast (M'_\ast) \; dM'_\ast = \int_{\mathcal{H}}^{\infty} \phi_{\rm h} (\mathcal{H}') \; d \mathcal{H}'.
    \label{eq:sham}
\end{equation}
Similar arguments hold true if we instead analyze subpopulations, such as centrals and satellite galaxies (see e.g., Rodr{\'\i}guez-Puebla et al. 2012), between star-forming/blue and quiescent/red galaxies (see e.g., Rodr{\'\i}guez-Puebla et al. 2011) or even other galaxy property that scales in a monotonic way with halos (see e.g., Shankar et al. 2006).

Another form to express Eq. \ref{eq:sham} is as follows: 
\begin{equation}
    \phi_\ast \left(M_\ast (\mathcal{H}) \right) = \phi_{\rm h} \left(\mathcal{H}\right) \frac{d\mathcal{H}}{d M_\ast(\mathcal{H})},
    \label{eq:dif_sham}
\end{equation}
which indicates that the GSMF is proportional to the halo function, except for a factor related to the slope $d\mathcal{H}/d M_\ast(\mathcal{H})$. In the specific case where $\mathcal{H} = M_{\rm peak}$, the above equation is related to the slope of the SHMR.

It is interesting to note that if the SHMR is independent of time, the SFE discussed in Section \ref{sec:galaxy_halos} is simply proportional to the slope of the SHMR:
\begin{equation}
    {\rm SFE}(M_{\rm peak}) =f_{\rm bar, U}^{-1} \cdot \frac{ d M_\ast( M_{\rm peak} ) }{d M_{\rm peak}}
\end{equation}
as shown in Rodr{\'\i}guez-Puebla et al. (2016). Then, the GSMF will be simply the halo mass function multiplied by the SFE:
\begin{equation}
    \phi_\ast \left(M_\ast (M_{\rm peak}) \right) = \phi_{\rm h} (M_{\rm peak})  \bigg/  f_{\rm bar, U} \cdot {\rm SFE}(M_{\rm peak}).
\end{equation}
where $f_{\rm bar, U}\sim 0.16$. Notice that if dark matter halos were 100\% efficient in transforming baryons to halos then ${\rm SFE}=1$ and therefore $M_\ast = f_{\rm bar, U} \cdot M_{\rm peak} $. Consequently, the GSMF will be given by $\phi_\ast (f_{\rm bar, U} \cdot M_{\rm peak}) = \phi_{M} (M_{\rm peak})  / f_{\rm bar, U} $.

\end{appendix}

\end{document}